# The Application of Stochastic Optimization Algorithms to the Design of a Fractional-order PID Controller

Mithun Chakraborty, Deepyaman Maiti, and Amit Konar
Department of Electronics and Telecommunication Engineering
Jadavpur University
Kolkata, India
mithun.chakra108@gmail.com, deepyamanmaiti@gmail.com, konaramit@yahoo.co.in

*Abstract*—The Proportional-Integral-Derivative Controller is widely used in industries for process control applications. Fractional-order PID controllers are known to outperform their integer-order counterparts. In this paper, we propose a new technique of fractional-order PID controller synthesis based on peak overshoot and rise-time specifications. Our approach is to construct an objective function, the optimization of which yields a possible solution to the design problem. This objective function is optimized using two popular bio-inspired stochastic search algorithms, namely Particle Swarm Optimization and Differential Evolution. With the help of a suitable example, the superiority of the designed fractional-order PID controller to an integer-order PID controller is affirmed and a comparative study of the efficacy of the two above algorithms in solving the optimization problem is also presented.

*Keywords-Differential evolution; dominant poles; integer-order and fractional-order PID controllers; particle Swarm Optimization*

I. INTRODUCTION

The merit of using a Proportional-Integral-Derivative (PID) controller lies in its simplicity of design and good performance, including low percentage overshoot and small settling time (which is essential for slow industrial processes). PID controllers belong to the class of dominating industrial controllers and, therefore, continuous efforts are being made to improve their quality and robustness. An elegant way of enhancing the performance of PID controllers is to use *fractional-order controllers* where the I- and D-actions have, in general, non-integer orders.

In order to grasp the significance of *fractional-order PID controllers*, an understanding of the *theory of fractional calculus* is necessary. Fractional calculus is that branch of mathematical analysis [13], which generalizes the order of the derivative or integral of a function to a *real number* (not necessarily an integer). If $D$ denotes first-order differentiation, then, we know $D^2$ denotes two iterations of differentiation. Likewise, $D^{1/2}$ may be interpreted as some operator which, when applied twice to a function successively, will have the same effect as a single differentiation [13]. Similar explanations hold for fractional integration too. Just as first-order differentiation (or integration) of a function in time-domain maps to multiplication by $s^1$ (or $s^{-1}$) of the Laplace Transform of the function in *s*-domain, $s^\alpha$ indicates time-domain derivation to the order $\alpha$ if $\alpha > 0$ or time-domain integration to the order $|\alpha|$ if $\alpha < 0$. The name given to this generalized differential/integral operation is *differintegration*. Of the several definitions of *fractional differintegrals*, the Grünwald-Letnikov and Riemann-Liouville definitions [14] are the most used. These definitions are required for the realization of discrete control algorithms.

In a fractional PID controller, besides the proportional, integral and derivative constants, denoted by $K_p$, $T_i$ and $T_d$ respectively, we have two more adjustable parameters: the powers of s in integral and derivative actions, $-\lambda$ and $\delta$ respectively. As such, this type of controller has a wider scope of design, while retaining the advantages of classical PID controllers. Finding the appropriate settings of the values of the five parameters $\{K_p, T_i, T_d, \lambda, \delta\}$ to achieve optimal system performance thus calls for optimization on the five-dimensional space. Classical optimization techniques are not applicable here because of the roughness of the multidimensional objective function surface. We, therefore, use derivative-free optimization techniques: the first one — Particle Swarm Optimization (PSO) — draws inspiration from the intelligent, collective behavior of a swarm of social insects (particularly bees) foraging for food together and the other — Differential Evolution (DE) — is an evolutionary algorithm that is guided by the principles of Darwinian Evolution and Natural Genetics [12].

Traces of work on fractional-order PID controllers are available in the current literature [1]-[9] on control engineering. A frequency domain approach based on the expected crossover frequency and phase margin is mentioned in [2]. A method based on pole distribution of the characteristic equation in the complex plane was proposed in [5]. A state-space design method based on feedback poles placement can be viewed in [6]. The fractional-order controller can also be designed by cascading a proper fractional unit to an integer-order controller.

Our design focuses on positioning closed loop dominant poles, and finding the optimal set of values of the design parameters that satisfy the constraints thus obtained on the characteristic equation. The work is thus original and may open up new avenues for the next generation fractional-order controller design. Moreover, as it is already proven that the performance of fractional-order PID controllers surpasses that of the classical ones with integro-differential operations of integer orders [3], our proposed design is likely to find extensive applications in real industrial processes.

## II. THE INTEGER AND FRACTIONAL ORDER PID CONTROLLERS

The integer-order PID controller has the following transfer function: $K_p + T_i s^{-1} + T_d s$. Here, the orders of integration and derivation are both unity. The real objects or *processes* that we wish to control are generally fractional in order (for example, the voltage-current relation of a semi-infinite lossy RC line). As, for many of them the fractionality is very low, integer-order approximations are applied. In general, however, the integer-order approximation of the fractional systems can cause significant differences between the mathematical model and the real system. The main reason for using integer-order models was the absence of solution methods for fractional-order differential equations.

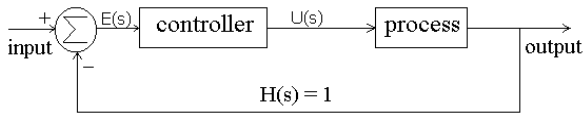

Figure 1. Block diagram of a unity-feedback closed loop control system

A fractional PID controller has the transfer function:

$$K_p + T_i s^{-\lambda} + T_d s^{\delta},$$

where λ and δ are positive real numbers. Taking λ =1, δ =1, we will have an integer-order PID controller. Thus we see that, while the integer-order PID controller has three parameters, its fractional-order counterpart has as many as five.

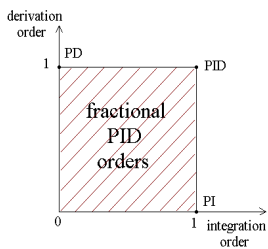

Figure 2. Expanding from point to plane

The fractional-order PID controller *expands* the integer-order PID controller *from point to plane*, as shown in Fig. 2., thereby adding flexibility to controller design and allowing us to control our real world processes more accurately.

## III. REVIEW ON PSO AND DE ALGORITHMS

### A. The Optimization Problem

The optimization problem consists in determining the *global optimum* (in our case, minimum) of a continuous real-valued function of *n* independent variables $x_1, x_2, x_3, \ldots, x_n$, mathematically represented as $f(\vec{X})$, where $\vec{X} = (x_1, x_2, x_3, \ldots, x_n)$ is called the *parameter vector*. Then the task of any optimization algorithm reduces to searching the *n*-dimensional hyperspace to locate a particular point with position-vector $\vec{X}_0$ such that $f(\vec{X}_0)$ is the global optimum of $f(\vec{X})$.

### B. Particle Swarm Optimization

PSO [10], [11], [12] developed by Eberhart & Kennedy, is in principle a multi-agent parallel search technique. We begin with a *population* or *swarm* consisting of a convenient number, say *m*, of *particles* — conceptual entities that "fly" through the multi-dimensional search space as the algorithm progresses through discrete (unit) time-steps $t = 0, 1, 2, \ldots$, the population-size *m* remaining constant.

In the standard PSO algorithm, each particle *P* has two state variables: its current position $\vec{X}_i(t) = [X_{i,1}(t), X_{i,2}(t), \ldots, X_{i,n}(t)]$ and its current velocity $\vec{V}_i(t) = [V_{i,1}(t), V_{i,2}(t), \ldots, V_{i,n}(t)]$, $i=1,2,\ldots,m$. The position vector of each particle with respect to the origin of the search space represents a *candidate solution* of the search problem. Each particle also has a small memory comprising its *personal best* position experienced so far, denoted by $\vec{p}_i(t)$ and the *global best* position found so far, denoted by $\vec{g}(t)$. Here, one position is considered *better* than another if the former gives a *lower* value of the objective function, also called the *fitness function* in this context, than the latter.

For each particle, each component $X_{i,j}(0)$ of the initial position vector is selected at random from a predetermined search range $[X_j^L, X_j^U]$, while each velocity component is initialized by choosing at random from the interval $[-V_{j\max}, V_{j\max}]$, where $V_{j\max}$ is the maximum possible velocity of any particle in the *j*th dimension, $j = 1, 2, \ldots, n$, $i = 1, 2, \ldots, m$; the initial settings for $\vec{p}_i(t)$ and $\vec{g}(t)$ are taken as $\vec{p}_i(0) = \vec{X}_i(0), \vec{g}(0) = \vec{X}_k(0)$ such that $f(\vec{X}_k(0)) \leq f(\vec{X}_i(0)) \forall i$.

After the particles are initialized, the iterative optimization process begins, where the positions and velocities of all the particles are updated by the following recursive equations (1) and (2). The equations are presented for the *j*th dimension of the position and velocity of the *i*th particle.

$$V_{i,j}(t+1) = \omega V_{i,j}(t) + C_1 \varphi_1 \cdot (p_{i,j}(t) - X_{i,j}(t)) + C_2 \varphi_2 \cdot (g_j(t) - X_{i,j}(t)) \quad (1)$$

$$X_{i,j}(t+1) = X_{i,j}(t) + V_{i,j}(t+1) \quad (2)$$

where the algorithmic parameters are defined as :

ω : inertial weight factor,
$C_1, C_2$ : two constant multipliers called *self confidence* and *swarm confidence* respectively,
$\varphi_1, \varphi_2$ : two uniformly distributed random numbers.

We take $V_{j\max} = X_j^U - X_j^L \ \forall j$, $\omega = 0.729$, $C_1 = C_2 = 1.494$, $0 < \varphi_1, \varphi_2 \leq 1$.

### C. Differential Evolution

DE [12], [15], [17] belongs to the class of *evolutionary algorithms* where each time-varying parameter vector (candidate solution) in the population is called a *chromosome* and each time-step represents a *generation*. The first step of the algorithm, as usual, is:

*Initialization*: This step is identical to the random initialization of position vectors in PSO.

Each iteration consists of the following three steps:

*Mutation*: For each chromosome $\vec{X}_i(t)$ belonging to the current generation, three other chromosomes $\vec{X}_p(t)$, $\vec{X}_q(t)$, and $\vec{X}_r(t)$ are randomly selected from the same generation (*i*, *p*, *q* and *r* are distinct); the *scaled difference* of $\vec{X}_q(t)$ and $\vec{X}_r(t)$ is added to $\vec{X}_p(t)$ to generate a *donor vector* $\vec{V}_i(t+1)$:

$$\vec{V}_i(t+1) = \vec{X}_p(t) + \mathbf{F} \cdot (\vec{X}_q(t) - \vec{X}_r(t))$$

where **F** is a *constant scalar* belonging to (0,1). We take **F** = 0.8.

*Recombination*: In this step, a *trial offspring vector* $\vec{T}_i(t+1)$ is created for each current-generation *parent vector* $\vec{X}_i(t)$ by first choosing a constant *CR* (0<*CR*<1) called the *crossover constant* and then setting the *j*th component $T_{i,j}(t+1)$ of $\vec{T}_i(t+1)$ according to the following criterion:

$$T_{i,j}(t+1) = \begin{cases} V_{i,j}(t) & \text{if } rand_j(0,1) < CR, \\ X_{i,j}(t) & \text{otherwise,} \end{cases}$$

where $rand_j(0, 1)$ is a random number selected from the interval (0, 1), $j = 1, 2,..., n$. We take $CR = 0.96$.

*Selection:* This step is guided by the principle of "survival of the fittest" and may be mathematically expressed as follows:

$$\vec{X}_i(t+1) = \begin{cases} \vec{T}_i(t+1) & \text{if } f(\vec{T}_i(t+1)) < f(\vec{X}_i(t)), \\ \vec{X}_i(t) & \text{otherwise,} \end{cases}$$

$\forall i = 1, 2, ......., m$.

Thus, the next-generation population is generated, keeping the population-size *m* always unchanged.

### IV. FORMULATION OF THE OBJECTIVE FUNCTION AND ITS OPTIMIZATION

Our approach is based on the *root locus* method (*dominant poles* method) of designing integral PID controllers [16].

As in the traditional root locus method, the peak overshoot $M_p$ and rise time $t_{rise}$ (or, in other words, requirements of stability and damping levels) are specified. From these specifications, we find out the damping ratio $\zeta$ and the undamped natural frequency $\omega_n$, making use of the following formulae [16]:

$$\zeta = -\frac{ln(M_p)}{\sqrt{\{ln(M_p)\}^2 + \pi^2}} \quad (3)$$

$$\omega_n = \frac{\pi - tan^{-1}\left(\frac{\sqrt{1-\zeta^2}}{\zeta}\right)}{t_{rise}\sqrt{1-\zeta^2}} \quad (4)$$

Using these computed values of $\zeta$ and $\omega_n$, we then determine the desired positions of the dominant poles $p_{1,2}$ of the closed loop system [16]:

$$p_{1,2} = -\zeta\omega_n \pm \mathbf{j}\omega_n\sqrt{1-\zeta^2} \quad (5)$$

$$= -a \pm \mathbf{j}\,b, \text{ where } a = \zeta\omega_n, b = \omega_n\sqrt{1-\zeta^2},$$

for $a, b > 0$.

Let $G_p(s)$ be the transfer function of the process we want to control, $G_c(s)$ the transfer function of the controller to be designed and $H(s)$ the transfer function of the feedback-path, as shown in Fig. 1.

Then, the closed loop transfer function of the controlled system is:

$$T(s) = \frac{G(s)}{1 + G(s)H(s)},$$

where $G(s) = G_c(s).G_p(s)$ is the forward path transfer function. As we use fractional-, and $G_c(s)$ is of the form

$$G_c(s) = K_p + T_i s^{-\lambda} + T_d s^\delta \quad (6)$$

Therefore, in general, the characteristic equation of the closed-loop system is:

$$1 + G(s)H(s) = 0 \quad (7)$$

Assuming unity feedback, we have $H(s)=1$. In this case, the characteristic equation becomes

$$1 + G(s) = 0$$

or, $1 + G_c(s).G_p(s) = 0$

or, $1 + G_c(s).P(s)/Q(s) = 0$

or, $Q(s) + G_c(s).P(s) = 0 \quad (8)$

where $P(s)$ and $Q(s)$ are respectively the numerator and denominator polynomials of $G(s)$, $P(s)$ and $Q(s)$ have no common factor.

As $p_{1,2}$ must be poles of the closed loop system, each of them must be a root of the characteristic equation (7) and hence must satisfy equation (8). Thus, putting $s = p_1 = -a \pm \mathbf{j}\,b$ in equation (8), we obtain

$$Q(p_1) + G_c(p_1).P(p_1) = 0$$

or, $Q(-a+\mathbf{j}b)+[K_p+T_i(-a+\mathbf{j}b)^{-\lambda}+T_d(-a+\mathbf{j}b)^{\delta}].P(-a+\mathbf{j}b) = 0$ (9)

Equation (9) is a complex equation in five unknowns, namely $K_p$, $T_i$, $T_d$, $\lambda$, $\delta$ and our problem of designing a controller, which makes the closed loop dominant poles of the system coincide with $p_{1,2}$, now reduces to determining the set of values of $\{K_p, T_i, T_d, \lambda, \delta\}$ for which (9) holds good. But as the number of unknowns exceeds the number of equations, there exists an infinite number of solution sets and the equation cannot be unambiguously solved by traditional methods.

This necessitates the application of stochastic global search techniques, which, in turn, requires the formulation of a suitable *objective function* or *cost function*.

Let

$R$ = real part of the L.H.S. of equation (9),

$I$ = imaginary part of the L.H.S. of equation (9),

$P = \tan^{-1}(I/R)$.

We define $f(K_p, T_i, T_d, \lambda, \delta) = |R| + |I| + |P|$ as our objective function.

Clearly, $f \geq 0$, in general, and $f = 0$ if and only if $R = 0$ and $I = 0$ and $P = 0$, i.e. if and only if equation (9) is satisfied.

So, we now employ first PSO and then DE to scour the five-dimensional search space $K_p \geq 0$, $T_i \geq 0$, $T_d \geq 0$, $0 \leq \lambda \leq 2$, $0 \leq \delta \leq 2$ and home in on the optimal solution set $\{K_p^*, T_i^*, T_d^*, \lambda^*, \delta^*\}$ for which $f = f_{min} = 0$.

The limits on the components of the position-vectors of the particles/chromosomes (i.e. the controller parameters) are set by us as follows: as a practical consideration, we assume

$0 \leq K_p, T_i, T_d \leq 1000$, $\quad 0 \leq \lambda, \delta \leq 2$.

In order to design an integer order PID controller for controlling the same process, we simply set $\lambda = \delta = 1$, so that the solution space becomes three-dimensional. All other conditions remain the same and the optimization process is executed as before to obtain the optimal set $\{K_p^*, T_i^*, T_d^*\}$.

It is germane to mention here that the formulae (3), (4) and (5) hold strictly only for a *second order system* with its *complex conjugate pole-pair* at $p_{1,2}$. However, for a higher-order system with a *dominant pole-pair* (real part of this pole-pair is much smaller than those of other poles), these formulae are widely used in control engineering applications with a fair degree of accuracy [16]. Nevertheless, after designing our controller with the help of these formulae, we perform a simulation to obtain the *unit step response* [16] of the closed-loop control system, as a check.

## V. ILLUSTRATION

### A. Problem Statement

The process (control objective) has the transfer function

$$G_p(s) = \frac{1}{0.8s^{2.2} + 0.5s^{0.9} + 1}.$$

We want to design a controller such that the closed loop system has a peak overshoot $M_p \leq 10\%$ and rise-time $t_{rise} \leq 0.3$ seconds.

### B. Solution

Using the formulae (3) and (4), for the limiting case, we obtain $\zeta = 0.5912$ and $\omega_n = 9.107$ s$^{-1}$. Thus the dominant poles for the closed loop controlled system should lie at $p_1 = (-5.384 + \mathbf{j}7.345)$ and $p_2 = (-5.384 - \mathbf{j}7.345)$.

As usual, ee assume unity feedback. The controller transfer function is given by (6). Putting $s = p_1 = (-5.384+\mathbf{j}7.345)$ in the characteristic equation, we obtain:

$$1 + \frac{K_p + T_i(-5.384 + \mathbf{j}7.345)^{-\lambda} + T_d(-5.384 + \mathbf{j}7.345)^{\delta}}{0.8(-5.384 + \mathbf{j}7.345)^{2.2} + 0.5(-5.384 + \mathbf{j}7.345)^{0.9} + 1} = 0$$

$$\Rightarrow [K_p + 13.4235 + \frac{T_i}{9.107^{\lambda}}\cos(2.203\lambda) + T_d(9.107)^{\delta}\cos(2.203\delta)]$$

$$+ \mathbf{j}[-98.9237 - \frac{T_i}{9.107^{\lambda}}\sin(2.203\lambda) + T_d(9.107)^{\delta}\sin(2.203\delta)] = 0.$$

(10)

After separating the real and imaginary parts, we have:

$$R = (K_p + 13.4235) + \frac{T_i}{9.107^{\lambda}}\cos(2.203\lambda) + T_d(9.107)^{\delta}\cos(2.203\delta)$$

(11)

$$I = -\frac{T_i}{9.107^{\lambda}}\sin(2.203\lambda) + T_d(9.107)^{\delta}\sin(2.203\delta) - 98.9237 \quad (12)$$

$$P = \tan^{-1}(I/R). \quad (13)$$

Thus, (11), (12) and (13) give us our objective function

$f(K_p, T_i, T_d, \lambda, \delta) = |R| + |I| + |P|$ which is minimized by PSO and by DE separately.

If we set $\lambda = 1$ and $\delta = 1$ before running the optimization algorithm, we obtain the three optimized parameters for the integer order PID controller. All results are presented in the next section.

Although we have constructed the objective function $f$ by making use of the by making use of dominant pole $p_1 = -a+\mathbf{j}b$ in the second quadrant, we would arrive at the same $f$ if we had started with the third-quadrant dominant pole $p_2 = -a-\mathbf{j}b$. This is because, in the latter case, the imaginary part of the reduced characteristic equation (9) would just be the negative of what we have obtained in (10) so that *f*, which involves absolute values only, would remain unaltered. This is true not only for the particular problem in question but in general as well.

TABLE I. RESULTS OF OPTIMIZATION FOR FRACTIONAL ORDER PID CONTROLLER DESIGN

| Algo. used | Optimized parameters for fractional-order PID controller | | | | |
|---|---|---|---|---|---|
| | $K_p$ | $T_i$ | $T_d$ | $\lambda$ | $\delta$ |
| PSO | 419.57 | 638.72 | 49.83 | 0.25 | 1.26 |
| DE | 962.80 | 197.55 | 46.27 | 1.79 | 1.37 |

TABLE II. RESULTS OF OPTIMIZATION FOR INTEGER ORDER PID CONTROLLER DESIGN

| Algo. used | Optimized parameters for integer-order PID controller ($\lambda=\delta=1$) | | |
|---|---|---|---|
| | $K_p$ | $T_i$ | $T_d$ |
| PSO | 60.86 | 14.03 | 13.63 |
| DE | 59.20 | 1.23 | 13.48 |

## VI. RESULTS

Although we allowed a maximum of 5000 iterations of PSO, we found that, in about 1000-1500 iterations, the *fitness value* (i.e. value of *f*) of the best particle dropped below the *tolerance* value of 0.0001 (practically equal to the perfect value of zero) and almost all other particles had fitness values very close to the best. Similar observations were made for optimization by DE.

Table I gives us the values of the controller parameters obtained using PSO and DE for fractional order while Table II gives us the corresponding data for the integer-order case.

From these tables, the expressions for the controller transfer function are obtained as shown in Table III:

TABLE III. CONTROLLER TRANSFER FUNCTIONS

| Algo. used | Order of PID controller $G_c(s)$ | |
|---|---|---|
| | Fractional | Integral |
| PSO | $419.57 + 638.72 s^{-0.25} + 49.83\, s^{1.26}$ | $60.86 + 14.03 s^{-1} + 13.63\, s$ |
| DE | $962.80 + 197.55 s^{-1.79} + 46.27\, s^{1.37}$ | $59.20 + 1.23 s^{-1} + 13.48 s$ |

With each of these expressions for $G_c(s)$, we compute the overall system transfer function $T(s)$ and obtain the time-response of the corresponding system to a unit step input $R(s)=1/s$ by finding the inverse Laplace Transform of $T(s)/s$.

Fig. 3. shows the plots of the unit step responses of the uncontrolled open-loop system given in the example considered, the closed-loop system controlled by the integral PID controller designed using PSO and the same system controlled by the fractional PID controller also designed using PSO. Fig. 4. shows the corresponding plots for the design using DE.

The values of $M_p$ and $t_{rise}$ calculated graphically for each of the controlled systems designed are presented in Table IV.

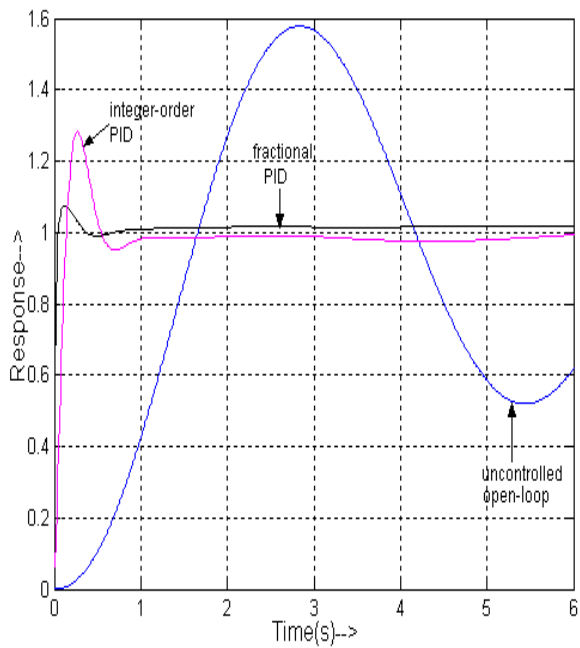

Figure 3. Unit Step Responses for design using PSO

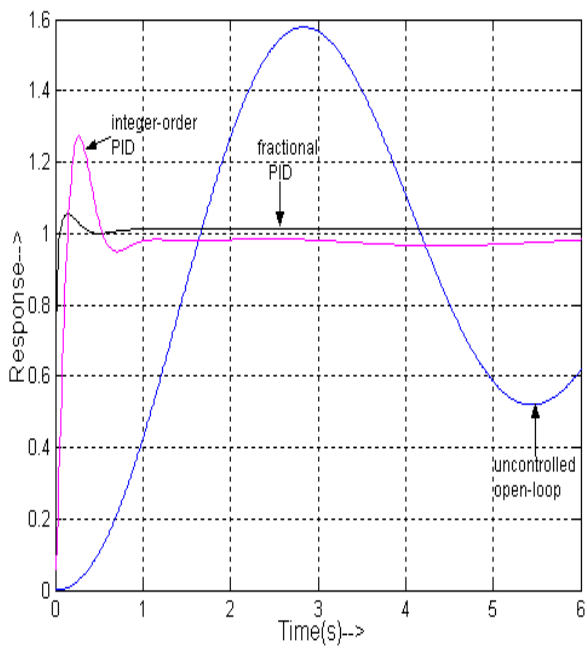

Figure 4. Unit Step Responses for design using DE

TABLE IV. PEAK OVERSHOOT AND RISE-TIME DATA FOR THE DIFFERENT SYSTEMS DESIGNED

| Order of controller | $M_p$ (%) | | $t_{rise}$ (seconds) | |
|---|---|---|---|---|
| | PSO | DE | PSO | DE |
| integral | 28.0 | 27.3 | 0.045 | 0.140 |
| fractional | 7.5 | 5.2 | 0.040 | 0.054 |

VII. CONCLUSION AND FUTURE RESEARCH DIRECTIONS

From Table III, many facts are apparent. Firstly, the rise-time requirement is satisfactorily met by all controllers whereas the integer-order controllers perform poorly in terms of peak overshoot. By comparing the rows of the table, we can easily observe the significant reduction in both $M_p$ and $t_{rise}$ (which translates to improved system performance or better compliance with a given set of user specifications) that may be achieved by replacing the traditional integer-order PID controller with its fractional counterpart. Again, a careful comparison of the columns of the table reveals that the overall results (particularly with respect to $M_p$) given by DE are notably better than those given by PSO.

Apart from the example shown, we also considered several other similar problems, solved them by our proposed method and drew similar inferences from the results thereof. These have not been included in this paper owing to paucity of space.

Hence, we may conclude that our design method is viable as it does produce fractional PID controllers superior to the ones with integral orders; moreover, DE appears to be a better option than PSO as an algorithm for minimizing the relevant objective function.

An important point is to be noted in this context. For the final system to exhibit the desired performance, it is necessary that the evaluated pole-values $p_{1,2}$ truly correspond to the *dominant poles* of the closed-loop controlled system. However, the optimization of our cost function ensures only that $p_{1,2}$ are poles of the system but does not guarantee that these are the dominant poles. In other words, equation (9) embodies a *necessary*, but not *sufficient*, condition for the dominance of $p_{1,2}$. Obviously, there may be a number of possible combinations of values of the parameters [$K_p$, $T_i$, $T_d$, .. ] for which the resulting characteristic equation is satisfied by $p_{1,2}$ but, in each such case, the characteristic equation will have other roots (closed-loop system poles), too, and these poles will also play an important role in determining the overall system response. So, using each of PSO and DE, we solved the optimization problem several times and naturally did not obtain the same result every time. The time response for each solution set was studied and the best one was reported. This is also a possible explanation of the observation that the performance of the final designed system with fractional controller is actually better than desired.

In constructing our objective function, we took the coefficient of each of the terms |R|, |I| and |P| to be unity, which means that we attached equal weights to each of these terms. Intuitively, it is understood that the first two terms are more important than the third. We are currently investigating what improvement in our method could be achieved by varying the coefficients of these terms or by constructing a superior cost function (such as one that also checks whether the evaluated poles are truly dominant). We also intend to apply other well-known algorithms such as Genetic Algorithm and its many variants to the same optimization problem and compare their performance with that of the two already studied.